\documentclass[pra,twocolumn,superscriptaddress,notitlepage,showpacs,showkeys]{revtex4-1}
\usepackage{graphicx,amsmath,amsfonts,amssymb,upgreek,txfonts,color}
\usepackage[colorlinks,linkcolor=blue,citecolor=blue,urlcolor=blue,breaklinks=true]{hyperref}

\begin{document}

\title{Force sensing in hybrid Bose-Einstein condensate optomechanics based on parametric amplification
}


\author{Ali Motazedifard} 
\email{motazedifard.ali@gmail.com}
\address{Department of Physics, Faculty of Science, University of Isfahan, Hezar Jerib, 81746-73441, Isfahan, Iran}
\address{Quantum Optics Group, Department of Physics, Faculty of Science, University of Isfahan, Hezar Jerib, 81746-73441, Isfahan, Iran}

\author{A. Dalafi} 
\email{a\_dalafi@sbu.ac.ir}
\address{Laser and Plasma Research Institute, Shahid Beheshti University, Tehran 19839-69411, Iran}

\author{F. Bemani}
\email{foroudbemani@gmail.com}
\address{Department of Physics, Faculty of Science, University of Isfahan, Hezar Jerib, 81746-73441, Isfahan, Iran}

\author{M. H. Naderi} 
\email{mhnaderi@sci.ui.ac.ir}
\address{Department of Physics, Faculty of Science, University of Isfahan, Hezar Jerib, 81746-73441, Isfahan, Iran}\address{Quantum Optics Group, Department of Physics, Faculty of Science, University of Isfahan, Hezar Jerib, 81746-73441, Isfahan, Iran}


\date{\today}
\begin{abstract}

In this paper, the scheme of a force sensor is proposed which has been composed of a hybrid optomechanical cavity containing an interacting cigar-shaped Bose-Einstein condensate (BEC) where the \textit{s}-wave scattering frequency of the BEC atoms as well as the spring coefficient of the cavity moving end-mirror (the mechanical oscillator) are parametrically modulated. It is shown that in the red-detuned regime and under the so-called impedance-matching condition, the mechanical response of the system to the input signal is enhanced substantially which leads to the amplification of the weak input signal while the added noises of measurement (backaction noises) can be suppressed and lowered much below the standard quantum limit (SQL). Furthermore, because of its large mechanical gain, such a modulated hybrid system is a much better amplifier in comparison to the (modulated) bare optomechanical system which can generate a stronger output signal while keeping the sensitivity nearly the same as that of the (modulated) bare one. The other advantages of the presented nonlinear hybrid system accompanied with the mechanical and atomic modulations in comparison to the bare optomechanical cavities are its controllability as well as the extension of amplification bandwidth.

\end{abstract}


\maketitle

\section{Introduction}
As is well-known, every measurement, at either the classical or quantum level, is affected by noise which reduces the accuracy of the measurement. Therefore, finding methods, especially in quantum systems, for noise suppression, noise cancellation, or signal amplification is of particular interest and importance in quantum measurements and quantum metrology.
For example, the so-called coherent quantum noise cancellation (CQNC) scheme has been recently introduced \cite{CQNCPRL,CQNCPRX} in which the ``anti-noise" path in the quantum dynamics of the system can be employed to cancel the original noise path via destructive quantum interference.

During the past decade, optomechanical systems (OMSs), in which the electromagnetic radiation pressure is coupled to a mechanical oscillator (MO) as a macroscopic object, have been developed \cite{Aspelmeyer,milburnoptomechanics,chenoptomechanics,Meystreoptomechanics} for the purpose of testing the fundamentals of physics like the Bell test \citep{optomechanicalBelltest1,*optomechanicalBelltest2} and emergence of quantum effects in macro scale.
Also, OMSs have been applied to a wide variety of research fields including ultra precision force sensing \cite{xsensing1}, MO ground-state cooling \cite{groundstatecooling,Sidebandcooling,Lasercooling}, generation of bipartite entanglement \cite{Palomaki2,Paternostro,genesentangelment}, synchronization of MOs \cite{Mari1synch,MianZhang,Bagheri,grebogi,foroudsynch}, generation of mechanical/optical nonclassical states \cite{Borkje,Hammerer,foroudnonlinearcoherent,unconditionalstate,addecoherentstate}, quantum simulation of the parametric dynamical Casimir effect (DCE) \cite{DCE1,DCEPRL,DCEQWell,aliDCE1,aliDCE2,aliDCE3} as well as the curved space-time \cite{foroudCurvedspacetime}, and generation of squeezing \cite{Clerkdissipativeoptomechanics,gentlymodulating,pontinmodulation,clerkfeedback,Harris,asjad,twofoldsqueezing}.


In optomechanical force sensors, the competition between the shot noise and the radiation pressure backaction noise which have opposite dependence on the input power, determines the standard quantum limit (SQL). In fact, increasing the input power makes the shot noise decrease, but nevertheless it causes an increase of the backaction noise. Therefore, in order to improve the force sensing precision one has to find methods to suppress or evade the backaction noise \cite{Sillanpaahiddencorrelation2018,Sillanpaanoiselessmeasurement2017}.

There are different theoretical and experimental proposals for backaction noise reduction to overcome the SQL in ultra precision force measurements \cite{khaliliforcesensing,clerkfeedback,experimentforce1,pontinexperimentforce3,teufelexperimentforce4}. In addition to these proposals which are based on noise reduction, the CQNC proposals are based on the noise cancellation via quantum interference \cite{CQNCmeystre,CQNCmaximilian,aliNJP,complexCQNC,PolzikCQNCnature,PolzikCQNCPRL}.
It should be noted that although in these methods the backaction noise of measurement is reduced or even canceled but the signal is not amplified at all. 
In a more recent proposal \cite{optomechanicswithtwophonondriving} it has been shown that it is possible to suppress the added noise of measurement while amplifying the input signal simultaneously in a bare optomechanical system through the parametric modulation of the spring coefficient of the MO.

On the other hand, recently proposed hybrid optomechanical cavities containing Bose-Einstein condensates (BECs) \cite{MeystreBEC,BrennBECexp,RitterBECexp} in which the fluctuation of the collective excitation of the BEC, i.e., Bogoliubov mode, behaves like an effective mechanical mode \cite{MeystreBEC} and the nonlinear atom-atom interaction simulates an atomic amplifier \cite{dalafi1,dalafi3}, have more controllability and can increase the quantum effects at macroscopic level \cite{dalafi2,dalafi6,dalafi7,dalafi8,BhattacherjeeNMS}. Besides, such hybrid systems are suitable for reduction of quantum noise \cite{Bhattacherjeenoisereduction} or can act as a quantum amplifier/squeezer \cite{aliDCEsqueezing}. Moreover, by considering the quadratic optomechanical coupling in such hybrid systems, one can generate robust entanglement and strong mechanical squeezing beyond the SQL \cite{dalafiQOC}.  

Based on the theory of linear quantum amplifiers \cite{forcedetection2}, in order to enhance the functionality of a linear amplifier it is necessary to add more degrees of freedom to the system so that the input signal is amplified more effectively. However, the price to pay for introducing extra degrees of freedom will be the manifestation of some added noises to the input signal. Nevertheless, there exist methods which can reduce the added noises of measurement below the SQL. For example, it has been recently shown \cite{Sillanpaabackaction2016} experimentally that in an optomechanical system with two mechanical modes one can achieve a measurement precision below the SQL based on a back-action evasion method.


Here, inspired by the above-mentioned investigations on optomechanical force sensors and the properties of the hybrid OMSs, we propose an experimentally feasible scheme for the weak force measurement beyond the SQL based on simultaneous signal amplification and backaction noise suppression via parametric amplification of the mechanical and Bogoliubov modes. 
We consider a hybrid optomechanical cavity with a moving end-mirror in the red-detuned regime containing an interacting cigar-shaped BEC in the dispersive regime of atom-field interaction where the \textit{s}-wave scattering frequency of the BEC atoms as well as the spring coefficient of the cavity moving end-mirror (the MO) are parametrically modulated.

The most important advantage of using a BEC in an optomechanical cavity is that the Bogoliubov mode of the BEC behaves effectively as a mechanical oscillator (a moving mirror) with a controllable natural frequency while the ordinary mechanical oscillators (moving end mirrors or membranes) have fixed natural frequencies which cannot be changed after fabrication. Therefore, as was mentioned previously, in order to enhance the functionality of a quantum amplifier, we need to use more extra degrees of freedom and since the BEC has more controllability we have chosen it as another extra mode. 

It is shown that because of its large mechanical gain such a hybrid system with both the atomic and mechanical modulations is a much better amplifier in comparison to the (modulated) bare optomechanical system which can generate a stronger output signal while keeping the sensitivity nearly the same as that of the (modulated) bare one studied in Ref.~\cite{optomechanicswithtwophonondriving}. Furthermore, the force measurement precision in the off-resonance region can be improved in such hybrid systems through the increase of the amplification bandwidth.


The paper is organized as follows. In Sec.~\ref{sec2} the system Hamiltonian is described and then in Sec.~\ref{sec.dynamics} the quantum Langevin-Heisenberg equations of motion are derived. In Sec.~\ref{sec.sensing}, it is shown how the input signal can be amplified by the enhancement of the mechanical response of the system while suppressing the added noise of measurement via parametric modulations of the mechanical and atomic modes. In Sec.\ref{Sen and SNR} the system sensitivity as well as the signal-to-noise ratio (SNR) are calculated and it is shown how the presented theoretical predictions can be realized in an experimental setup .Finally, the summary and conclusions are mentioned in Sec.~\ref{summary}.


\section{THEORETICAL DESCRIPTION OF THE SYSTEM}\label{sec2}

\begin{figure}
	\includegraphics[width=8.5cm]{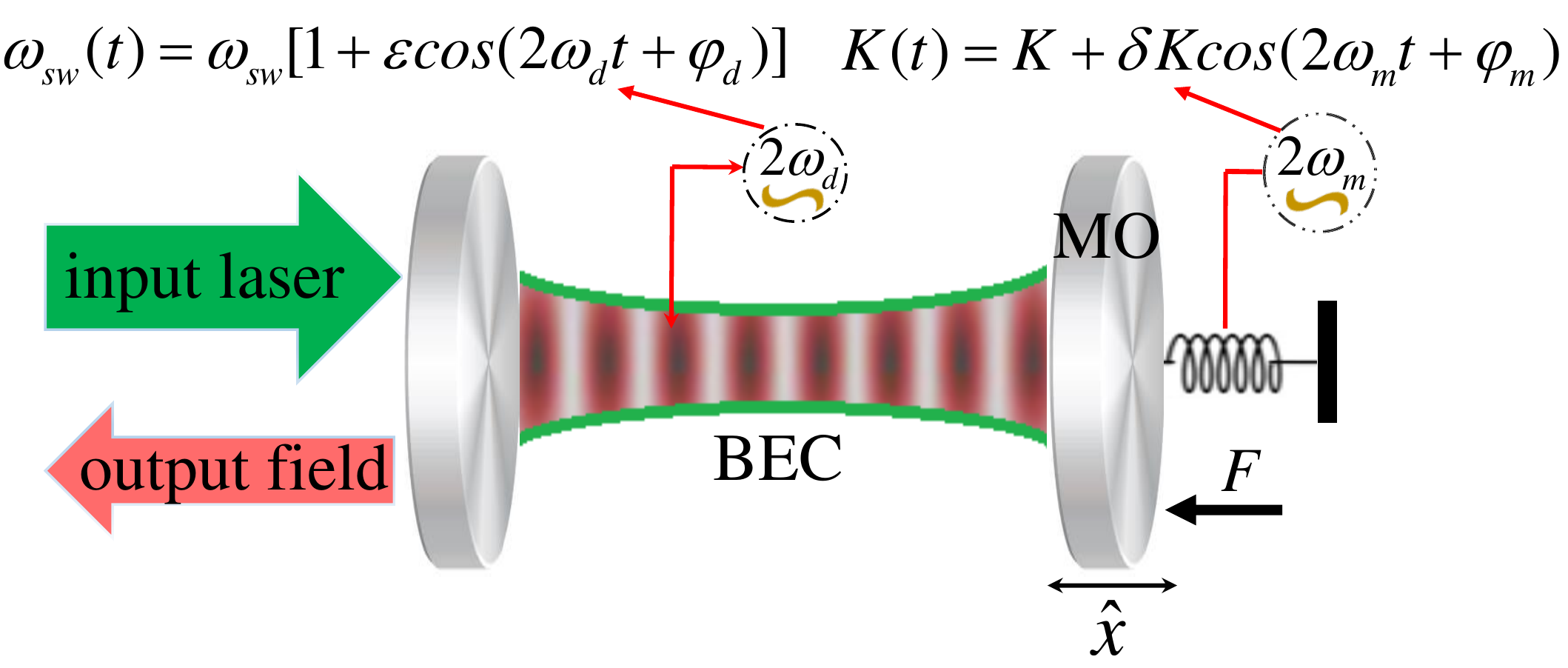}
	\caption{(Color online) Schematic of a hybrid optomechanical force sensor designed for the measurement of a weak force signal ($F$) exerting to the moving end-mirror of an optomechanical cavity which contains a BEC where the \textit{s}-wave scattering frequency of the condensate atoms as well as the spring coefficient of the cavity moving mirror are parametrically modulated, i.e., $ \omega_{sw}(t)=\omega_{sw}[1+\varepsilon \cos(2\omega_d t+\varphi_d)] $ and $ K(t)=K + \delta K \cos (2\omega_m t +\varphi_m) $. Also, the cavity mode is coherently driven by a classical laser field through the fixed mirror.}
	\label{fig1}
\end{figure}

As depicted in Fig.~(\ref{fig1}), the force sensor we are going to investigate is an optomechanical cavity with length $ L $ and damping rate $\kappa$ having a moving end-mirror with mass $m$, natural frequency $\omega_m$, and damping rate $ \gamma_m $ whose spring coefficient is parametrically modulated at twice its natural frequency. The cavity which is driven through the fixed mirror by a laser with frequency $\omega_{L}$ and wavenumber $k_{0}=\omega_{L}/c$ contains a BEC of $ N $ ultracold two-level atoms with mass $ m_a $ and transition frequency $ \omega_a $. Furthermore, we assume that the collision frequency of the condensate atoms is parametrically modulated through the modulation of the electromagnetic trap or the density of the BEC by changing the trap stiffness \cite{JaskulaBECmodulationexp}. The total Hamiltonian of the system in the frame rotating at the driving laser frequency $ \omega_L $ can be written as

\begin{eqnarray} \label{H1total}
&& \hat H= \hbar \Delta_c \hat a^\dag \hat a + i\hbar E_L (\hat a^\dag - \hat a)+\hbar \omega_m \hat b^\dag \hat b-\hbar g_0 \hat a^\dag \hat a (\hat b + \hat b^\dag)\nonumber\\
&& +\frac{i \hbar}{2} (\lambda_m \hat b^{\dag 2}  e^{-2i\omega_m t}- \lambda_m^\ast \hat b^2 e^{2i\omega_m t}) + \hat H_F + \hat H_{BEC} .
\end{eqnarray}
The first three terms in the Hamiltonian describe, respectively, the free energy of the cavity mode, the coupling
between the cavity mode and the driving laser, and the free energy of the MO. Here, $ \Delta_c= \omega_c-\omega_L $ is the detuning of the optical mode from the driving laser frequency, $E_L$ is the pump rate of the external laser, and $ \hat a $ ($ \hat b $) is the annihilation operator of the cavity (MO) mode. The canonical position and momentum of the MO are $ \hat x= x_{zp}(\hat b + \hat b^\dag) $ and $ \hat p= \hbar (\hat b - \hat b^\dag)/2ix_{zp} $, respectively, with $ x_{zp}= \sqrt{\hbar / 2m\omega_m} $ being the zero-point position fluctuation. 

The fourth term in Hamiltonian (\ref{H1total}) is the optomechanical interaction between the mechanical and optical modes with the single-photon optomechanical coupling $ g_0=x_{zp} \omega_c /L $. The fifth term describes the parametric driving of the MO spring coefficient at twice its natural frequency [$ K(t)=K+\delta K \cos (2\omega_m t + \varphi_m) $ with $ \varphi_m $ being the phase of external modulation] which is written in the rotating wave approximation (RWA) over time scales longer than $ \omega_m^{-1} $ where $ \lambda_m= \vert \lambda_m \vert e^{i\varphi_m} $ with $ \vert \lambda_m \vert= \delta K x_{zp}^2 / 2\hbar $ \cite{aliDCE3,optomechanicswithtwophonondriving}. Note that by fixing the phase of modulation $ \varphi_m $, it is always possible to take $ \lambda_m $ as a real number. It is worth to point out that this term can be considered as the mechanical phonon analog of the degenerate parametric amplification (DPA) which may lead to the DCE of mechanical phonons \cite{aliDCE3}.
The sixth term, $ \hat H_F $, accounts for the coupling of the MO to the input classical-force $ F $ to be measured which is given by 
\begin{eqnarray} \label{H_F}
&& \hat H_F=F(t) \hat x= x_{zp} F(t) (\hat b + \hat b^\dag).
\end{eqnarray}

The last term of Eq.~(\ref{H1total}) is the Hamiltonian of the atomic BEC. If the atom-laser detuning $\Delta_a=\omega_{a}-\omega_{L}$ is much greater than the atomic linewidth, then the excited electronic state of the atoms can be adiabatically eliminated and the Hamiltonian of the BEC can be written as \cite{Ritsch}

\begin{eqnarray} \label{H1_BEC}
&&\!\!\!\!\!\!\!\!\!\!\!\!\!\!\! \hat  H_{BEC}=\int_{-L/2}^{L/2}dx \hat \psi^\dag(x) \Big[\frac{-\hbar^2}{2m_a} \frac{d^2}{dx^2} \nonumber\\ 
&& + \hbar U_0 \cos^2(k_0 x) \hat a^\dag \hat a+\frac{1}{2}U_{s} \hat \psi^{\dag}(x)  \hat \psi^{\dag}(x)\Big] \hat \psi(x),
\end{eqnarray}
where $ \hat \psi(x) $ is the quantum field operator of the atomic BEC, $ U_0=-g_a^2/\Delta_a $ is the optical lattice barrier height per photon, $ g_a $ is the atom-field coupling constant, $ U_s=4\pi \hbar^2 a_s/m_a $, and $ a_s $ is the two-body \textit{s}-wave scattering length \cite{Ritsch,Domokos2}. For the weak atom-field interaction, the quantum field operator of the BEC under the Bogoliubov approximation can be expanded as\cite{Nagy}
\begin{eqnarray} \label{si1}
&& \hat \psi(x)= \sqrt{N/L}+ \sqrt{2/L}\cos(2k_0x) \hat d ,
\end{eqnarray}
where $ \hat d $ is the Bogoliubov mode of the BEC which corresponds to the quantum fluctuations of the atomic field around the classical condensate mode $ \sqrt{N/L} $. By substituting Eq.~(\ref{si1}) into Eq.~(\ref{H1_BEC}) the Hamiltonian of the BEC is obtained as follows
\begin{eqnarray}   
&&\!\!\!\!\!\!\!\!  \hat H_{BEC} =  \hbar\delta_0 \hat a^\dag \hat a +  \hbar \omega_d \hat d^\dag \hat d + \hbar G_0\hat a^\dag \hat a (\hat d  + \hat d^\dag)  + \hat H_{sw}  , \label{H_BEC}
\end{eqnarray}	
where $ \delta_0=NU_0/2 $, $ \omega_d= 4 \omega_R + \omega_{sw} $ is the effective frequency of the Bogoliubov mode of the BEC, $ \omega_R=\hbar k_0^2/2m_a $ is the recoil frequency of the condensate atoms, and $ G_0=\sqrt{2N}U_0/4 $ is the radiation pressure coupling between the Bogoliubov mode of the BEC and the optical mode. 

The Hamiltonian $ \hat H_{\textit{sw}}$ in Eq.~(\ref{H_BEC}) refers to the atom-atom interaction energy. In the presence of time modulation of the \textit{s}-wave scattering frequency of atomic collisions at twice the frequency of the Bogoliubov mode, i.e., $ \omega_{sw}(t)= \omega_{sw}[1+\varepsilon \cos (2\omega_d t+\varphi_d)]$ where $ \varepsilon $ and $ \varphi_d $ are, respectively, the amplitude and the phase of modulation, and $ \omega_{sw}= 8\pi\hbar N a_s/(m_a L w^2) $ with $ w $ being the beam waist of the optical mode, $ \hat H_{sw} $ in the RWA is given by
\begin{eqnarray} \label{H_sw_mod}
&& \hat H_{sw}(t) = \frac{i \hbar}{2} (\lambda_d \hat d^{\dag 2}  e^{-2i\omega_d t}- \lambda_d^\ast \hat d^2 e^{2i\omega_d t}),
\end{eqnarray}
where $ \lambda_{d}=-i\varepsilon \omega_{sw} e^{-i\varphi_d}/4 $ can be taken real by fixing the phase $ \varphi_d $.

It should be noted that the \textit{s}-wave scattering frequency, $ \omega_{sw} $, can be controlled experimentally by manipulating the transverse trapping frequency of the BEC through changing the waist radius of the optical mode $ w $ \cite{morsch}. Besides, as has been shown in Ref.~\cite{JaskulaBECmodulationexp} the time modulation of the atomic collisions can be experimentally realized by the time modulation of the scattering length via the modulation of the electromagnetic trap, or the modulation of the density of the BEC by changing the trap stiffness via the intensity modulation of the pump laser.

The Hamiltonian of Eq.~(\ref{H_sw_mod}) is a Bogoliubov-phonon analog of the DPA which can give rise to the generation of Bogoliubov-type Casimir phonons \cite{aliDCE3}. Here, it should be mentioned that in the Hamiltonian (\ref{H_BEC}), we have ignored the cross-Kerr nonlinear coupling between the intracavity field and the Bogoliubov mode which is negligibly small in comparison to the radiation pressure interaction \cite{dalafi7}.

Substituting Eqs.~(\ref{H_F}), (\ref{H_BEC}), and (\ref{H_sw_mod}) into the Hamiltonian of Eq.~(\ref{H1total}) the total Hamiltonian of the system takes the form
\begin{eqnarray}\label{Hamt}
&& \hat H_{tot}=\hbar \Delta_0 \hat a^\dag \hat a + \hbar \omega_m \hat b^\dag \hat b + \hbar \omega_d \hat d^\dag \hat d + i \hbar E_L(\hat a^\dag - \hat a ) \nonumber \\
&& \qquad   -\hbar g_0 \hat a^\dag \hat a (\hat b+\hat b^\dag) + \hbar G_{0} \hat a^\dag \hat a (\hat d +\hat d^\dag)+x_{zp} F(t) (\hat b + \hat b^\dag)  \nonumber \\
&& \qquad + i \frac{\hbar }{2} (\lambda_m\hat b^{\dag2} e^{-2i\omega_m t}- \lambda_m^\ast \hat b^2 e^{2i\omega_m t})\nonumber\\
&&\qquad+i\frac{ \hbar}{2} (\lambda_d \hat d^{\dag 2}  e^{-2i\omega_d t}- \lambda_d^\ast \hat d^2 e^{2i\omega_d t}), 
\end{eqnarray}
where $ \Delta_0=\Delta_c + NU_0/2 $ is the cavity Stark-shifted detuning.

\section{Dynamics of the system \label{sec.dynamics}}
The linearized quantum Langevin equations (QLEs) of the system can be derived from the Hamiltonian of Eq.~(\ref{Hamt}). As has been shown in Ref.~\cite{aliDCE3} in the red detuned regime and within the RWA where the two optomechanical and opto-atomic couplings are analogous to the beam-splitter interaction, the linearized QLEs describing the dynamics of the quantum fluctuations are given by
\begin{subequations} \label{QLEBS}
	\begin{eqnarray}
	&& \!\!\!\!\! \delta \dot { \hat a} = -\frac{\kappa}{2} \delta \hat a + i g  \delta \hat b - i G \delta \hat d + \sqrt{\kappa} \hat a_{in} , \\
	&&\!\!\!\!\! \delta \dot {\hat b}  = -\frac{\gamma_m}{2} \delta \hat b + i g \delta \hat a + \lambda_m \delta \hat b^\dag -i\frac{x_{zp}}{\hbar} F e^{i\omega_m t}+  \sqrt{\gamma_m}\hat b_{in}, \\
	&&\!\!\!\!\! \delta \dot {\hat d}  = - \dfrac{\gamma_d}{2} \delta \hat d  -i G  \delta \hat a + \lambda_d \delta \hat d^\dag  +  \sqrt{\gamma_d} \hat d_{in},
	\end{eqnarray}
\end{subequations}
where $ g=g_0 \bar a $ and $ G = G_0 \bar a $ are, respectively, the enhanced-optomechanical and opto-atomic coupling strengths in which $\bar a=E_L/\sqrt{\kappa^2/4+\bar \Delta_{0}^{2}} $ is the steady-state mean value of the optical mode. Here, $ \bar \Delta_{0}=\Delta_{0}-2g_0 \bar b+2G_0 \bar d $ is the effective cavity detuning where $ \bar b\approx g_0 \bar a^2/\omega_{m}  $ and $ \bar d\approx -G_0 \bar a^2/\omega_{d} $ are, respectively, the steady-state values of the mechanical and atomic mean fields in the RWA and in the high quality factors limit. Besides, $\gamma_{m}$ and $\gamma_{d}$ are the dissipation rates of the mechanical and the Bogoliubov modes, respectively. 

Here, the red-detuned regime of cavity optomechanics is defined by the condition $ \bar\Delta_0 \approx \omega_m \approx \omega_d $. For this purpose, the frequency of the Bogoliubov mode of the BEC, i.e., $ \omega_d $, should be matched to the mechanical frequency ($ \omega_d \approx \omega_m $) which is possible through the manipulation of the \textit{s}-wave scattering frequency of the Bogoliubov mode via controlling the transverse frequency of the BEC trap \cite{morsch}. Besides, the effective detuning $\bar\Delta_0$ can be set in the red-detuning regime through the pump laser frequency.

Furthermore, the optical input vacuum noise $ \hat a_{in} $ as well as the Brownian noises $ \hat b_{in} $ and $ \hat d_{in} $ affecting, respectively, the MO and the Bogoliubov mode of the BEC, satisfy the Markovian correlation functions $ \langle \hat a_{in}(t) \hat a_{in}^\dag (t') \rangle= (1+\bar n_c^T) \delta (t-t') $, $ \langle \hat a_{in}^\dag (t) \hat a_{in}(t') \rangle= \bar n_c^T \delta (t-t') $; $ \langle \hat o_{in}(t) \hat o_{in}^\dag (t') \rangle= (1+\bar n_j^T) \delta (t-t') $ and $ \langle \hat o_{in}^\dag (t) \hat o_{in}(t') \rangle= \bar n_j^T \delta (t-t') $ with $ o= b $  and $  d $ where $ \bar n_j^T=[exp(\hbar \omega_j /k_B T)-1]^{-1} $ with $ j=c, m $ and $ d $ are the mean number of thermal excitations of the cavity, mechanical, and Bogoliubov modes at temperature $ T $. The quantum noise $\hat d_{in}$ originates from the other extra modes of the BEC as well as the fluctuations in the electromagnetic trap as has been shown in Ref.~\cite{dalafi4}.

Now by defining the quadratures $ \delta \hat X_o=(\hat o+\hat o^{\dagger})/\sqrt{2} $ and $ \delta \hat P_o=(\hat o-\hat o^{\dagger})/\sqrt{2}i $ ($ o=a,b,d $) the set of Eqs.~(\ref{QLEBS}a)-(\ref{QLEBS}c) can be written as the following compact matrix form 
\begin{eqnarray} \label{udot}
&&\delta \dot {\hat u}(t)= A ~\delta \hat u(t) + \hat u_{in}(t),
\end{eqnarray}
where the vector of continuous-variable fluctuation operators and the corresponding vector of noises are, respectively, given by $ \delta\hat u=\Big(\delta \hat X_a,\delta \hat P_a,\delta \hat X_b, \delta \hat P_b,\delta \hat X_d,\delta \hat P_d\Big)^T $ and $ \hat  u_{in}(t)=\Big(\sqrt{\kappa}\hat X_a^{in},\sqrt{\kappa}\hat P_a^{in},\sqrt{\gamma_m}\hat X_b^{\prime in},\sqrt{\gamma_m}\hat P_b^{\prime in},\sqrt{\gamma_d} \hat X_d^{in},\sqrt{\gamma_d} \hat P_d^{in}\Big)^T $ in which  $ \hat X_o^{in}=(\hat o_{in}+\hat o_{in}^{\dagger})/\sqrt{2} $ and $ \hat P_o^{in}=(\hat o_{in}-\hat o_{in}^{\dagger})/\sqrt{2}i $ ($ o=a,b,d $). Besides,
\begin{subequations} \label{md MN}
	\begin{eqnarray}
&&\hat X_b^{\prime in}(t)=\hat X_b^{in}(t)+\sqrt{\frac{2}{\gamma_m}}\frac{x_{zp}}{\hbar}F(t) \sin\omega_m t,\label{md1}\\
&&\hat P_b^{\prime in}(t)=\hat P_b^{in}(t)-\sqrt{\frac{2}{\gamma_m}}\frac{x_{zp}}{\hbar}F(t) \cos\omega_m t,\label{md2}
\end{eqnarray}
\end{subequations}
are the modified mechanical noises.
Furthermore, the time-independent drift matrix $ A $ is given by 
\begin{eqnarray} \label{A}
&&\!\!\!\!\!\!\!\!\!\!\!\!  A\!=\! \left( \begin{matrix}
{-\frac{\kappa}{2}} & {0} & {0} & {-g} & {0} & {G}  \\
{0} & {-\frac{\kappa}{2}} & {g} & {0} & {-G} & {0}  \\
{0} & {-g} & {\!\lambda_m\!-\!\frac{\gamma_m}{2}} & {0} & {0} & {0}  \\
{g} & {0} & {0} & {-(\lambda_m\!+\!\frac{\gamma_m}{2})} & {0} & {0}  \\
{0} & {G} & {0} & {0} & {\lambda_d -\frac{\gamma_d}{2}} & {0}  \\
{-G} & {0} & {0} & {0} & {0} & {-(\lambda_d\!+\!\frac{\gamma_d}{2} )}  \\
\end{matrix} \right).
\end{eqnarray}
As has been shown in Ref.~\cite{aliDCE3}, based on the Routh-Hurwitz criterion for the optomechanical stability condition, the parameters $ \lambda_{m} $ and $ \lambda_{d} $ should satisfy the condition
\begin{eqnarray} \label{stability}
&&\!\!\!\!\!\!\! \lambda_{m(d)} \le \! \frac{\gamma_{m(d)}}{2} \left[1+ \mathcal{C}_{m(d)}\right]:= \lambda_{m(d)}^{max},
\end{eqnarray}
in which $ \mathcal{C}_m $($ \mathcal{C}_d $) is the collective optomechanical cooperativity associated with the mechanical mode (Bogoliubov mode) given by 
\begin{eqnarray}
&&  \mathcal{C}_{m(d)}=  \mathcal{C}_{0(1)} \frac{1+ \mathcal{C}_{1(0)} - \xi_{d(m)}^2}{(1+ \mathcal{C}_{1(0)}-\xi_{d(m)}^2)^2- \xi_{d(m)}^2  \mathcal{C}_{1(0)}^2} ~,
\end{eqnarray}
where $\mathcal{C}_0=4g^2/\kappa \gamma_m$ and $\mathcal{C}_1= 4G^2/\kappa \gamma_d $ are the optomechanical and opto-atomic cooperativities, respectively, and $ \xi_{d(m)}=2\lambda_{d(m)}/\gamma_{d(m)}$ plays the role of an effective dimensionless-amplitude of modulation. 

The solution to the QLEs, i.e., Eq.~(\ref{udot}), in the Fourier space can be written as $ \delta\hat u(\omega)= \boldsymbol{  { \chi }}(\omega) \hat u_{in}(\omega) $ where $\boldsymbol{  { \chi }}(\omega)$ is the susceptibility matrix and the Fourier transforms of the modified mechanical noises, i.e., those of Eqs.(\ref{md1}) and (\ref{md2}) are as follows
\begin{subequations}
	\begin{eqnarray}
	&&\hat X_b^{\prime in}(\omega)=\hat X_b^{in}(\omega)-\sqrt{\frac{2}{\gamma_m}}\frac{x_{zp}}{\hbar}\frac{i}{2}[F(\omega+\omega_m)-F(\omega-\omega_m)],\nonumber\\
	&&\hat P_b^{\prime in}(\omega)=\hat P_b^{in}(\omega)-\sqrt{\frac{2}{\gamma_m}}\frac{x_{zp}}{\hbar}\frac{1}{2}[F(\omega+\omega_m)+F(\omega-\omega_m)].\nonumber
	\end{eqnarray}
\end{subequations}
 
Now, using the input-output theory for the field operators, the output P-quadrature of the cavity field, i.e., $\delta \hat P_a^{out}(\omega)=-\sqrt{\kappa} \delta\hat P_{a}(\omega)+ \hat P_{in}^{a}(\omega)  $, is obtained as follows

\begin{equation} \label{paoutomega}
\delta \hat P_a^{out}(\omega)= \mathcal{A}(\omega) \hat P_a^{in}(\omega)+ \mathcal{B}(\omega) \hat X_b^{\prime in}(\omega) + \mathcal{D}(\omega) \hat X_d^{in}(\omega) ,
\end{equation}
where $ \mathcal{A}(\omega)= 1-\kappa \chi_{22}(\omega) $, $ \mathcal{B}(\omega)=\sqrt{\kappa \gamma_{m}} \chi_{23}(\omega) $, and $ \mathcal{D}(\omega)=\sqrt{\kappa \gamma_{d}} \chi_{25}(\omega) $ and the relevant elements of the susceptibility matrix are given by
\begin{eqnarray}
&&\!\!\!\! \chi_{22}(\omega)=  \left[ \chi_0^{-1}(\omega)+g^2 \chi_{-m}+G^2 \chi_{-d}(\omega)  \right]^{-1} , \label{chi22} \nonumber \\
&&\!\!\!\! \chi_{23}(\omega)=g \left[ \chi_0^{-1}(\omega) \chi_{-m}^{-1}(\omega) \! +g^2 \! \! +G^2 \! \chi_{-d}(\omega)  \chi_{-m}^{-1}(\omega)   \right]^{-1} , \label{chi23} \nonumber \\
&&\!\!\!\! \chi_{25}(\omega)=-G \left[ \chi_0^{-1}(\omega) \chi_{-d}^{-1}(\omega) \! +G^2 \! \! +g^2 \! \chi_{-m}(\omega)  \chi_{-d}^{-1}(\omega)   \right]^{-1},  \label{chi25}\nonumber
\end{eqnarray}
with $ \chi_0^{-1}(\omega)= \kappa/2 - i\omega  $ and $ \chi_{- m(-d)}^{-1}(\omega)= \gamma_{m(d)}/2 - \lambda_{m(d)} - i \omega  $. It is clear that $ \chi_0(-\omega)= \chi_0(\omega)^\ast $ and $ \chi_{- m(-d)}(-\omega)= \chi_{- m(-d)}(\omega)^\ast $, so $ \chi_{ij}(-\omega)= \chi_{ij}(\omega)^\ast  $.

\section{single quadrature force sensing \label{sec.sensing}}
In this section by calculating the spectrum of the optical output phase quadrature, we will show how coherent modulations of both the atomic collisions frequency and the mechanical spring coefficient lead to the simultaneous signal amplification and backaction noise suppression which provides the best conditions for an ultra precision force sensing.

In the optomechanical force sensor demonstrated in Fig.~(\ref{fig1}), the imprint of the input mechanical signal is manifested in the  cavity output field through the optomechanical interaction. In other words, the MO position shift exerted by the external force leads to a change of the effective cavity length and therefore causes the variation of the optical cavity output phase. As a consequence, the signal corresponding to the exerted external force can be detected by measuring the spectrum of the optical output phase quadrature, $\hat P_a^{out}$, through methods like heterodyne, homodyne or synodyne detections \cite{complexCQNC}. 
In the following, we will show how the proposed hybrid optomechanical system allows us for single-quadrature force sensing with noise suppression and signal amplification which helps to surpass the SQL on force detection.
 
In order to measure and detect the input mechanical force, one should calculate the optical output phase quadrature spectrum, 
\begin{eqnarray}
S_{P_{a}}^{out}(\omega)&=&\frac{1}{4\pi}\int d\omega^{\prime}e^{i(\omega+\omega^{\prime})t} \langle \delta\hat P_{a}^{out}(\omega) \delta\hat P_{a}^{out}(\omega^{\prime}) \nonumber\\
&& + \delta\hat P_{a}^{out}(\omega^{\prime}) \delta\hat P_{a}^{out}(\omega) \rangle.
\end{eqnarray}
Since the signal has been coded in the input mechanical noise quadrature, for an efficient force-sensing, we should manipulate the system parameters such that the mechanical response to the input quadrature $ \hat X_b^{in} $ is amplified while the optical and atomic responses to the input noise quadratures $ \hat P_a^{in} $ and $ \hat X_d^{in} $ are attenuated. Since the mechanical response of the system is independent of the classical input signal force and depends only on the quantum properties of the system, in the following, we set aside the classical function $ F $ and calculate the optical output phase quadrature spectrum by considering just the input quantum noises. So the output optical power spectrum is obtained as
\begin{eqnarray}
&& S_{P_{a}}^{out}(\omega)= (\bar n_c^T+\frac{1}{2}) \vert \mathcal{A}(\omega) \vert^2 \nonumber \\ 
&& \qquad \qquad +(\bar n_m^T+\frac{1}{2}) \vert \mathcal{B}(\omega) \vert^2 \! + \! (\bar n_d^T+\frac{1}{2})  \vert \mathcal{D}(\omega) \vert^2. \label{s_pp_a} 
\end{eqnarray}
After some algebraic manipulations, one can rewrite the spectrum of the optical output phase quadrature as follows
\begin{eqnarray} \label{forcesensing1}
&& S_{P_{a}}^{out}(\omega)=R_m(\omega) \left[ (\bar n_m^T + \frac{1}{2})+n_{add}(\omega) \right],
\end{eqnarray}
where 
\begin{eqnarray}
&& R_m(\omega)=\vert \mathcal{B(\omega)} \vert^2=\kappa \gamma_m \vert \chi_{23}(\omega) \vert^2 \label{S_F}, \\
&& n_{add}(\omega)= (\bar n_c^T +\frac{1}{2}) \frac{\vert \mathcal{A}(\omega) \vert^2 }{\vert \mathcal{B}(\omega) \vert^2 }+ (\bar n_d^T + \frac{1}{2}) \frac{\vert \mathcal{D}(\omega) \vert^2}{\vert \mathcal{B}(\omega) \vert^2 } . \label{N_F}
\end{eqnarray}

Here, $ R_m(\omega) $ is the mechanical response to the input signal and $ n_{add}(\omega) $ is the added noise of measurement which originates from the contributions of the input optical and atomic vacuum noises to the phase quadrature of the output cavity field. As is seen from Eq.~(\ref{forcesensing1}), the added noise can be considered as an effective increase in the number of the thermal excitations of the mechanical reservoir due to the backaction of the optical and atomic modes. For a high precision force sensing and surpassing the SQL, one should amplify the mechanical response and suppress the added noise spectrum simultaneously. The SQL on force-sensing is defined as $ n_{add}^{\rm SQL}(\omega)=1/2 $ \cite{forcedetection1,forcedetection2} which has already been achieved experimentally \cite{SchrepplerSQLexperiment}. In the following it is shown that through the mechanical and atomic modulations the SQL can be surpassed by suppressing the added backaction noise especially near the on-resonance frequency of the output $ P $-quadrature while the input force signal is amplified through the enhancement of the system mechanical response.

The on-resonance added noise and mechanical response are, respectively, given by 
\begin{eqnarray} \label{signal and noise 1}
&&\!\!\!\!\!\!\!\!\!\!\!\!\! n_{add}(0)\!=\! \frac{(1-\xi_m)^2}{\mathcal{C}_0} \!\!  \left[\! \frac{\mathcal{G}_a}{(\sqrt{\mathcal{G}_a}\!-\!1)^2 } (\bar n_c^T\!+\!\frac{1}{2})\! +\! \frac{\mathcal{C}_1}{(1\!-\!\xi_d)^2} (\bar n_d^T\!+\!\frac{1}{2})\! \right], \label{N_F_0} \\
&&\!\!\!\!\!\!\!\!\!\!\!\!\!  R_m(0)=\mathcal{C}_0 \left(\frac{\sqrt{\mathcal{G}_a}-1}{1-\xi_m}\right)^2 , \label{S_F_0}
\end{eqnarray}
where the optical gain $ \mathcal{G}_a $, which is defined in the context of the linear amplifiers as the ratio of photons number in the output of the amplifier to that in the input \cite{forcedetection2}, is given by \cite{aliDCEsqueezing}
\begin{eqnarray}
&& \sqrt{\mathcal{G}_a}= \frac{\mathcal{C}_0-(1-\xi_m)+\mathcal{C}_1 \frac{1-\xi_m}{1-\xi_d}}{\mathcal{C}_0+(1-\xi_m)+\mathcal{C}_1 \frac{1-\xi_m}{1-\xi_d}} . \label{g_a}	
\end{eqnarray} 

As is seen from Eq.~(\ref{N_F_0}), the added noise is suppressed in the limit of $ \xi_m \to 1 $. However, as is seen from Eq.~(\ref{S_F_0}), in order to have signal amplification, the mechanical response should be increased simultaneously which is only possible when the optical gain is negligibly small or equal to zero, i.e., when $ \mathcal{G}_a= 0 $. To achieve zero gain, the impedance-matching condition given by
\begin{eqnarray} \label{impedance1}
&& \!\!\!\! \mathcal{C}_0+(\xi_m-1)[1-\mathcal{C}_1/(1-\xi_d)]=0, \quad    \mathcal{C}_0+ \mathcal{C}_1 \le 1 ,
\end{eqnarray}
should be satisfied. In other words, in order to have simultaneous noise suppression together with signal amplification, the numerical values of the cooperativities ($ \mathcal{C}_0 $ and $ \mathcal{C}_1 $) and also the atomic modulation $ \xi_{d} $ should be chosen so that the impedance-matching condition of Eq.(\ref{impedance1}) is satisfied for any specified value of the mechanical modulation in the limit of $\xi_m\to 1 $.

In the case where there is neither mechanical nor atomic modulation (off-modulations), i.e., $ \xi_{d}=\xi_m=0 $, we have
\begin{eqnarray}\label{modoff}
&& \!\!\!  n_{add}^{\rm {off}}(0)= \! \frac{1}{\mathcal{C}_0} \left[\frac{(\mathcal{C}_0+\mathcal{C}_1-1)^2}{4} (\bar n_c^T+\frac{1}{2}) + \mathcal{C}_1 (\bar n_d^T + \frac{1}{2}) \right] \! ,\label{noise_no_modulation}  \\
&& \!\!\! R_m^{\rm {off}}(0)=\frac{4\mathcal{C}_0}{(1+\mathcal{C}_0+\mathcal{C}_1)^2} \quad. \label{signal_no_modulation}
\end{eqnarray}
As is evident, in this case the mechanical response is always smaller than unity under the impedance-matching condition ($ \mathcal{C}_0+\mathcal{C}_1=1 $) while the added noise is fairly large. This means that in the off-modulations case, the system is able to transduce the mechanical force but cannot amplify the signal and suppress the added noise. In other words, it cannot operate as a high precision measurement device. 

In the other special case where there is no atomic modulation ($ \xi_d=0 $) while the mechanical modulation is turned on, the impedance-matching condition reads $ \xi_m + \mathcal{C}_0/(1-\mathcal{C}_1)=1 $, and consequently
\begin{eqnarray} \label{signal and noise 2}
&&  n_{add}(0)=\frac{\mathcal{C}_1}{\mathcal{C}_0} (1-\xi_m)^2 (\bar n_d^T +\frac{1}{2}), \label{N_F_1}  \\
&&  R_m(0)=\frac{\mathcal{C}_0 }{(1-\xi_m)^2}. \label{S_F_1}
\end{eqnarray}
In this case, it is clear that in the limit of $ \xi_m \to 1 $ there is a large mechanical response to the input signal with no added optical noise while there is a small residual backaction noise due to the Bogoliubov mode of the BEC.

\begin{figure}
	\includegraphics[width=8.6cm]{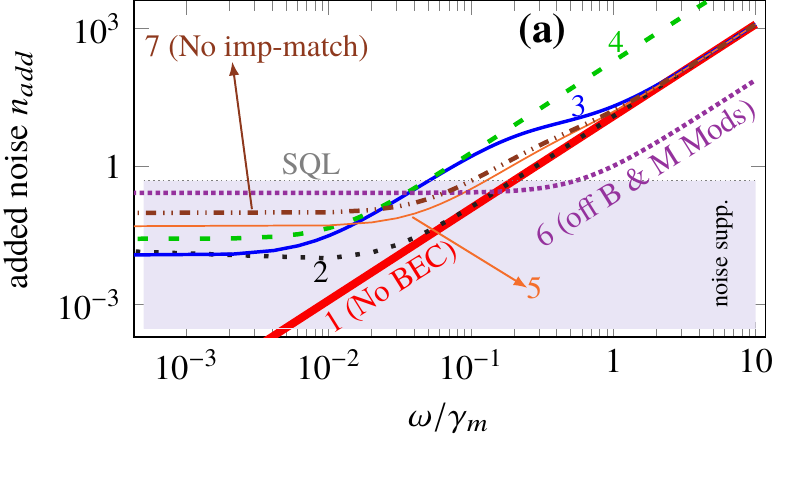}
	\includegraphics[width=8.6cm]{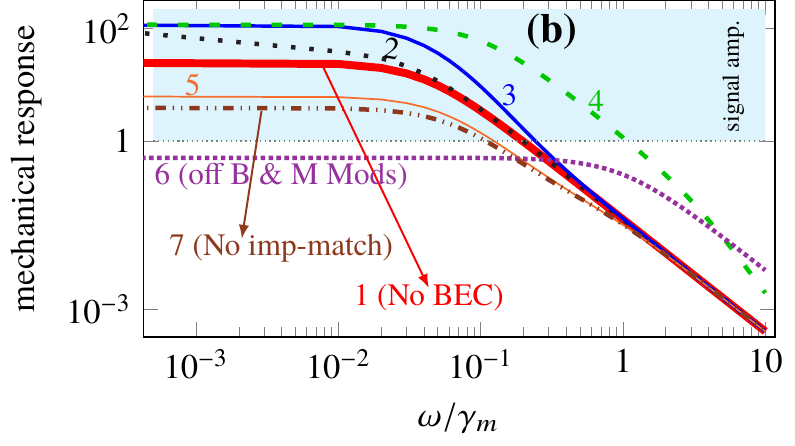}
	\caption{(Color online) (a) The added noise, $ n_{add}(\omega) $, and (b) the mechanical response to the input signal, $ R_m(\omega) $, vs dimensionless frequency $ \omega/\gamma_m $. The curves indicated by 1 to 5 and 7 have been plotted under the impedance-matching condition of Eq.~(\ref{impedance1}) with $ \mathcal{C}_0=0.04 $ and $ \mathcal{C}_1=0.5 $. The red solid very thick curve 1 corresponds to the absence of BEC with $ \xi_m=0.96 $, and $ \gamma_m/\gamma_d=1 $; the black dotted curve 2 corresponds to $ \xi_m=0.98,\xi_d=1.42 $, and $ \gamma_m/\gamma_d=10^2 $; the blue solid thick curve 3 corresponds to $ \xi_m=0.98,\xi_d=1.42 $, and $ \gamma_m/\gamma_d=1 $; the green dashed curve 4 corresponds to $ \xi_m=0.98, \xi_d=1.42 $ and $ \gamma_m/\gamma_d=10^{-2} $; the orange solid thin curve 5 corresponds to $ \xi_m=0.92, \xi_d=0 $, and $ \gamma_m/\gamma_d=1 $; the purple densely dotted curve 6 corresponds to $ \xi_m=0, \xi_d=0 $, and $ \gamma_m/\gamma_d=1 $, i.e., the off-modulations case with the impedance-matching condition $ \mathcal{C}_0+\mathcal{C}_1=1 $ ($ \mathcal{C}_0=\mathcal{C}_1=0.5 $); and the brown dashed-double-dotted curve 7 corresponds to the no impedance-matching condition with $ \xi_m=0.9, \xi_d=0.2 $, and $ \gamma_m/\gamma_d=1 $. The gray region (under the SQL line) in panel (a) and the cyan region in panel (b) correspond, respectively, to the situations where there are noise suppression and signal amplification. Here, we have assumed $ \kappa/\gamma_m\simeq 10^5 $, $ \bar n_m^T=10^3 $, and $ \bar n_c^T=\bar n_d^T \simeq 0 $. 
	}
	\label{fig2}
\end{figure}

In order to see how the mechanical and atomic modulations affect the signal amplification and noise suppression, in Fig.~(\ref{fig2}) we have plotted the added noise $ n_{add}(\omega) $ [Fig.~\ref{fig2}(a)] and the mechanical response to the signal $ R_m(\omega) $ [Fig.~\ref{fig2}(b)] versus the normalized frequency $\omega/\gamma_{m}$ in the largely different cooperativities
regime with $\mathcal{C}_0=0.04$ and $ \mathcal{C}_1=0.5 $ under the impedance-matching condition [curves indicated by 1 to 5]. Here, the effective modulation amplitudes, i.e., $\xi_{m}$ and $ \xi_d $, corresponding to the above specified values of cooperativities have been calculated based on the impedance-matching condition [Eq.~(\ref{impedance1})] together with the stability condition [Eq.~(\ref{stability})]. Besides, we have demonstrated the case of off-modulations ($\xi_m=0, \xi_d=0$) under the impedance-matching condition of $ \mathcal{C}_0+\mathcal{C}_1=1 $  with $\mathcal{C}_1=0.5$ [curve indicated by 6] and also the case of the absence of the impedance-matching condition [curve indicated by 7] for the sake of comparison with the other ones.

As is seen from Fig.~\ref{fig2}(b), in the  case of ``off-modulations" ($ \xi_m=\xi_d=0 $) [the densely dotted curve indicated by 6] the mechanical response is lower than unity which means that there is no signal amplification while in the presence of modulations [curves indicated by 1 to 5, and 7] the mechanical response gets larger than unity which leads to the signal amplification. Furthermore, the most efficient situation of signal amplification occurs under the impedance-matching condition. As is seen, the signal amplification in the absence of the impedance-matching condition [curve indicated by 7] is not as efficient as those under this condition.

On the other hand, in the absence of the BEC when the mechanical modulation is on [the red curve indicated by 1] which is similar to the situation studied in Ref.~\cite{optomechanicswithtwophonondriving}, there is a strong noise suppression together with a fairly good signal amplification notably near the on-resonance frequency ($ \omega \simeq 0 $). This shows how the presence of the mechanical modulation can lead to an ultra sensitive force measurement. However, the presence of the BEC together with atomic modulation improve the signal amplification substantially through the increase of the mechanical response of the system near the on-resonance frequency [compare the black, blue and green curves indicated, respectively, by 2, 3 and 4 with the red curve indicated by 1]. 

As a comparison, based on the results of Fig.\ref{fig2}(b), in the absence of the BEC the numerical value of the mechanical response near the resonance is $ R_{m}(\omega\approx 0)\approx 25 $ while in the presence of the BEC with atomic modulation this value is increased up to $ R_{m}(\omega\approx 0)\approx 118 $ which is much greater than that in the absence of the BEC. The physical reason for this increase is that the Bogoliubov mode of the BEC acts as an extra degree of freedom which based on the theory of linear amplifiers \cite{forcedetection2} leads to the enhancement of the amplifier functionality.

Naturally, the price paid for this strong improvement of the signal amplification is an increase in the added noise of the measurement because the presence of the BEC, as an extra phononic mode, induces an additional backaction noise. Nevertheless, the increment of the added noise due to the presence of the BEC is not so large to affect a precise measurement. As is seen from Fig.~\ref{fig2}(a), the added noise remains much below the SQL near the on-resonance frequency for the curves indicated by 2, 3 and 4.

Another advantage of the presence of the BEC with atomic modulation is the possibility of the ``off-resonance" force sensing. As is seen from Fig.~\ref{fig2}, the proposed optomechanical force sensor can amplify the signal [the cyan region in Fig.\ref{fig2} (b)] and attenuate the added noise [the gray region in Fig.\ref{fig2} (a)] in a wide range as large as $ \Delta \omega_{measurement} \sim \gamma_m/5 $ about the on-resonance frequency. In fact, by controlling the ratio of phononic damping rates such that $ \gamma_m/\gamma_d < 1 $ [see the green curve indicated by 4 for which $ \gamma_m/\gamma_d=0.01$] the signal can be amplified strongly in a much wider range around the off-resonance region (the bandwidth of amplification gets much larger). Therefore, the presence of the BEC together with atomic modulation improve the signal amplification effectively while in the absence of atomic modulation the BEC by itself does not enhance the signal amplification considerably [see the orange curves indicated by 5 in Fig.~(\ref{fig2})].

\begin{figure}
	\includegraphics[width=8.6cm]{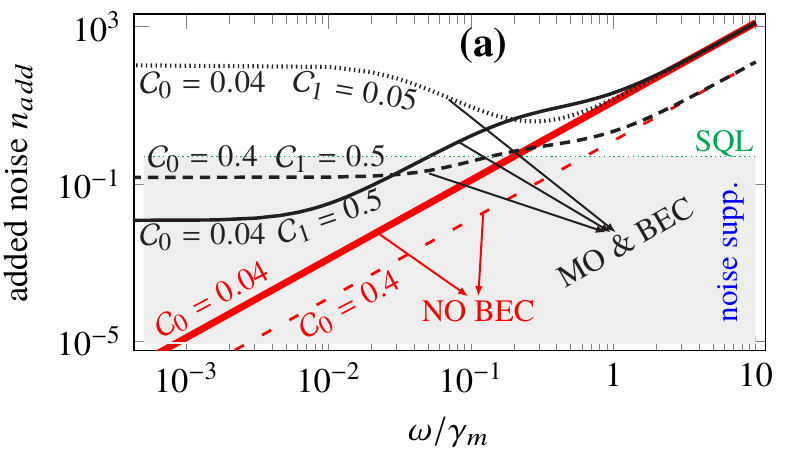}
	\includegraphics[width=8.6cm]{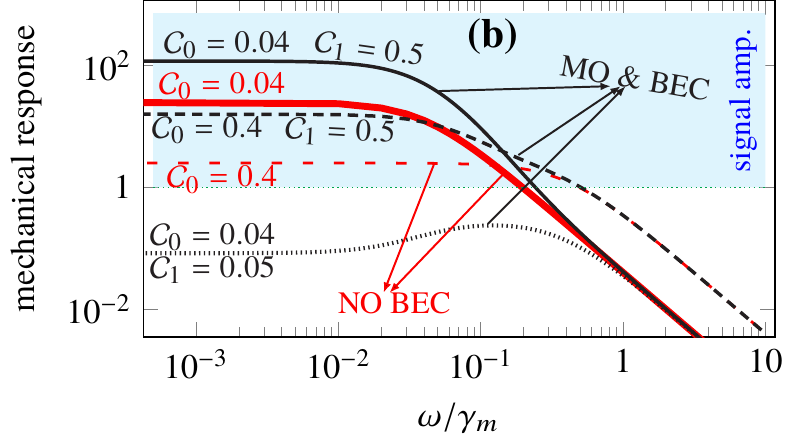}
	\caption{(Color online) (a) The added noise, $ n_{add}(\omega) $, and (b) the mechanical response to the input signal, $ R_m(\omega) $, vs dimensionless frequency $ \omega/\gamma_m $ for different ratios of cooperativities $ \mathcal{C}_{1}/\mathcal{C}_{0} $ under the impedance-matching condition of Eq.~(\ref{impedance1}). The red solid thick (indicated by $ \mathcal{C}_0=0.04 $) and the red loosely dashed (indicated by $ \mathcal{C}_0=0.4 $) curves correspond to the absence of BEC. The black solid thin (indicated by $ \mathcal{C}_0=0.04, \mathcal{C}_1=0.5 $), black densely dashed (indicated by $ \mathcal{C}_0=0.4, \mathcal{C}_1=0.5 $), and black dotted (indicated by $ \mathcal{C}_0=0.04, \mathcal{C}_1=0.05 $) curves correspond to the presence of the BEC when both atomic and mechanical modulations are turned on. The gray region (under the SQL line) in panel (a) and the cyan region in panel (b) correspond, respectively, to the situations where there are noise suppression and signal amplification. Here, we have set $ \gamma_m/\gamma_d=1 $. The other parameters are the same as those of Fig.~(\ref{fig2}).
	}
	\label{fig3}
\end{figure}

On the other hand, in order to see how the ratio of the atomic and the mechanical cooperativities affects the signal amplification and noise suppression, in Fig.~(\ref{fig3}) we have plotted the added noise [Fig.~\ref{fig3}(a)] and the mechanical response [Fig.~\ref{fig3}(b)] versus the normalized frequency $ \omega/\gamma_m $ for different ratios of cooperativities $ \mathcal{C}_1/\mathcal{C}_0 $ under the impedance-matching condition. For each curve represented in Fig.~(\ref{fig3}), the effective amplitudes of modulations ($\xi_{d}$ and $ \xi_m $) can be obtained from the impedance-matching condition (\ref{impedance1}) for the specified values of cooperativities. 
Here, the red solid thick and red loosely dashed curves indicated, respectively, by $ \mathcal{C}_0=0.04 $ [with $ \xi_m=0.96 $] and $ \mathcal{C}_0=0.4 $ [with $ \xi_m=0.6 $] correspond to the absence of the BEC, i.e., $\mathcal{C}_{1}=0$. Besides, the black solid thin curve indicated by $ \mathcal{C}_0=0.04, \mathcal{C}_1=0.5 $ [with $ \xi_m=0.98, \xi_d=1.42  $], the black densely dashed curve indicated by $ \mathcal{C}_0=0.4, \mathcal{C}_1=0.5 $ [with $ \xi_m=0.84, \xi_d=1.32 $], and the black dotted curve indicated by $ \mathcal{C}_0=0.04, \mathcal{C}_1=0.05 $ [with $ \xi_m=0.30, \xi_d=0.94 $] correspond to the presence of the BEC when both atomic and mechanical modulations are turned on.

As is seen clearly in Fig.~(\ref{fig3}), in the absence of the BEC an acceptable amount of signal amplification is achievable near the on-resonance frequency through the mechanical modulation for small values of mechanical cooperativities while the added noise is nearly equal to zero which is due to the absence of an extra mode (the red solid thick curve indicated by $ \mathcal{C}_0=0.04 $). Nevertheless, the presence of the BEC with a large ratio of $\mathcal{C}_1/\mathcal{C}_0$ together with both atomic and mechanical modulations lead to much stronger signal amplification while the added noise does not increase very much and stays much below the SQL (see the black solid thin curve indicated by $ \mathcal{C}_0=0.04, \mathcal{C}_1=0.5 $).

However, the signal amplification is reduced substantially by decreasing the ratio of $\mathcal{C}_1/\mathcal{C}_0$ (the black densely dashed curve indicated by $ \mathcal{C}_0=0.4, \mathcal{C}_1=0.5 $). Especially, for lower values of $\mathcal{C}_1$ (the black dotted curve indicated by $ \mathcal{C}_0=0.04, \mathcal{C}_1=0.05 $) not only there is no signal amplification (the signal is attenuated) but also the added noise increases significantly. Therefore, equipping the system with an extra atomic mode of a BEC together with atomic modulation can enhance the ability of signal amplification substantially while the extra added noise can be kept much below the SQL in a specific parametric regime which is based on the so-called impedance-matching condition with a large ratio of $\mathcal{C}_1/\mathcal{C}_0$.

Here, we have not taken into account the effect of the classical fluctuation in the phase of the external laser that drives the cavity, i.e., the so-called laser phase noise (LPN) in our theoretical model. Although the effect of the LPN is very important especially for large values of the laser linewidth in an ultra precision measurement but as has been shown in Ref.~\cite{mehmoodforcesensingdissipativelaserphasenoise2019}, as far as the laser linewidth is less than 1 kHz its effect becomes very negligible. In other words, for such low values of the laser line width, there is no necessity to take into account the effect of the LPN in a theoretical model.
 
Finally, it is worth to compare the presented method of force sensing which is based on parametric modulations with those based on the backaction-evasion \cite{clerkfeedback,forcedetection1} and CQNC techniques \cite{CQNCPRL,CQNCPRX,CQNCmeystre,aliNJP,PolzikCQNCnature}. The former is able to surpass the SQL by producing a large signal without suppressing the added noise while the latter can cancel the backaction noise completely without amplifying the signal. However, the force sensing scenario proposed in the present work, which is based on simultaneous signal amplification (mechanical response amplification) and noise suppression possesses the advantages of both the above-mentioned methods in that it provides a large signal to noise ratio. It should be pointed out that the improvement of force sensing in our scheme relies on small cooperativities with a large difference, i.e., $ \mathcal{C}_0+\mathcal{C}_1 < 1 $ and $ \mathcal{C}_0 \ll \mathcal{C}_1 $, which is achievable by taking a weak red-detuned driving together with the validity of the RWA.

\section{Sensitivity, Signal-to-noise ratio and experimental discussion}\label{Sen and SNR}
In this section, we calculate the system sensitivity  to the external force as well as the SNR so that the advantages of the present scheme in comparison to the others are clarified more explicitly. We also show how the presented theoretical predictions given in Figs.~(\ref{fig2}) and (\ref{fig3}) can be realized in an experimental setup.

Based on Eq.~(\ref{paoutomega}) and the explanations given in its previous paragraph, the output P-quadrature of the cavity field can be rewritten as 
\begin{equation}\label{dPa}
\delta\hat P_{a}^{out}(\omega)=\delta\hat P_{a}^{out}(\omega)|_{F=0}-\frac{i\mathcal{B}(\omega)}{\sqrt{m\hbar\omega_{m}\gamma_{m}}}\tilde{F}(\omega),
\end{equation}
where $ \delta\hat P_{a}^{out}(\omega)|_{F=0}=\mathcal{A}(\omega) \hat P_a^{in}(\omega)+ \mathcal{B}(\omega) \hat X_b^{in}(\omega) + \mathcal{D}(\omega) \hat X_d^{in}(\omega) $ is the contribution of the quantum noises and the second term is the transduction force with $ \tilde{F}(\omega)=\frac{1}{2}[F(\omega+\omega_{m})-F(\omega-\omega_{m})] $ being the external force. In order to calculate the sensitivity of the device to the external force and the SNR we define the force operator as \cite{jacobtaylorsensivity}
\begin{equation}\label{force operator}
\delta \hat F(\omega)=\frac{\delta\hat P_{a}^{out}(\omega)}{\partial\delta\hat P_{a}^{out}(\omega)/\partial\tilde F(\omega)}=\delta\hat N(\omega)+\tilde{F}(\omega),
\end{equation}
where
\begin{equation}
\delta\hat N(\omega)=i\frac{\sqrt{m\hbar\omega_{m}\gamma_{m}}}{\mathcal{B}(\omega)}\delta\hat P_{a}^{out}(\omega)|_{F=0}
\end{equation}
is the \textit{noise force} operator. The power spectrum of the noise force, i.e., $ S_N(\omega)=\frac{1}{2}\langle{\delta\hat N(\omega),\delta\hat N(\omega)^{\dagger}}\rangle $ is simply obtained as 
\begin{equation} \label{forcenoise}
S_{N}(\omega)=m\hbar\omega_{m}\gamma_{m}\Big[(\bar n^T_{m}+\frac{1}{2})+n_{add}(\omega)\Big].
\end{equation}
Based on the standard definition of the SNR \cite{bookstochastic}, it is the ratio of the signal, i.e., the absolute value of $\tilde{F}(\omega)$, to the variance of the noises, i.e., the square root of $S_N(\omega)$,
\begin{equation}\label{SNR}
r(\omega)=\frac{|\tilde{F}(\omega)|}{\sqrt{S_{N}(\omega)}}=\frac{|\tilde{F}(\omega)|}{\sqrt{m\hbar\omega_{m}\gamma_{m}}\sqrt{(\bar n^T_{m}+\frac{1}{2})+n_{add}(\omega)}} .
\end{equation}

The sensitivity or the minimum detectable input of the device is the minimum magnitude of the input signal required to produce an output with $r(\omega)=1$ \cite{jacobtaylorsensivity,vitaliSNR}. Therefore, $ \mathcal{S}(\omega)=\sqrt{S_N(\omega)} $  is obtained as
\begin{equation}\label{sensitivity}
\mathcal{S}(\omega)=\sqrt{m\hbar\omega_{m}\gamma_{m}}\Big[(\bar n^T_{m}+\frac{1}{2})+n_{add}(\omega)\Big]^{1/2}.
\end{equation}
As is seen from Eq.~(\ref{sensitivity}), the less the added noise the better the system sensitivity (especially for $n_{add}<1/2$ the SQL is surpassed).

In order to show how our theoretical predictions can be realized in an experimental setup and also for obtaining the numerical values of the sensitivity and the SNR we use the experimentally feasible parameters given in Refs.~\cite{BrennBECexp, RitterBECexp}.
For this purpose, we consider $ N=10^5 $ Rb atoms inside an optical cavity of length $ L=178 \mu$m with a damping rate of $ \kappa=2\pi\times 1.3 $MHz and the bare frequency $ \omega_{c}=2.41494\times 10^{15} $Hz corresponding to a wavelength of $ \lambda=780 $nm. The atomic $ D_{2} $ transition corresponding to the atomic transition frequency $ \omega_{a}=2.41419\times 10^{15} $Hz couples to the mentioned mode of the cavity. The atom-field coupling strength $ g_{a}=2\pi\times 14.1 $MHz and the recoil frequency of the atoms is $ \omega_{R}=23.7 $KHz. The movable end mirror can be assumed to have a mass of $ m=10^{-9}  $g and damping rate of $ \gamma_{m}=2\pi\times 100 $Hz which oscillates with frequency $ \omega_{m}=10^5 $Hz. In addition, the coherent modulation of the mechanical spring coefficient of the MO and also the time modulation of the \textit{s}-wave scattering frequency of atom-atom interaction of the BEC can be realized experimentally as have been reported, respectively, in \cite{pontinmodulation} and \cite{JaskulaBECmodulationexp}.

We now proceed to verify that the above-mentioned experimental data are compatible with the ranges of values of $ \mathcal{C}_0 $ and $ \mathcal{C}_1 $ studied in this paper. To do this, we show how the red-detuned regime of cavity optomechanics, i.e., the condition $ \bar\Delta_{0}=\omega_d=\omega_m $ is satisfied.
Firstly, the condition $ \omega_d=\omega_m $ determines the \textit{s}-wave scattering frequency as $ \omega_{sw}=\omega_{m}-4\omega_{R} $ which for $ \omega_{R}=23.7 $ kHz leads to $ \omega_{sw}=0.22\omega_{R} $ and consequently $ \omega_{d}=4.22\omega_{R} $. As has been explained previously the \textit{s}-wave scattering frequency is controllable experimentally through the transverse frequency of the electromagnetic trap of the BEC. 

Secondly, for the specified values of $ \mathcal{C}_0 $ and $ \mathcal{C}_1 $ the value of the atom-laser detuning is determined by
\begin{eqnarray}
&& \Delta_a=-\frac{g_{a}^{2}}{g_{0}}\sqrt{\frac{N \gamma_{m}}{8\gamma_{d}}\frac{\mathcal{C}_0}{\mathcal{C}_1}} .
\end{eqnarray}
 For example, for the the black solid curve of Fig.~(\ref{fig3}) (representing the presence of BEC with $ \mathcal{C}_0=0.04 $ and $ \mathcal{C}_1=0.5 $ together with the mechanical and atomic modulations) the atom-laser detuning is obtained as $ \Delta_a=-796.527 $ GHz which is an experimentally acceptable detuning to keep the system in the regime of dispersive atom-field interaction \cite{RitterBECexp}. In this way, the frequency of the external driving laser is determined by $ \omega_{L}=\omega_{a}-\Delta_a $ which for the black solid curve of Fig.\ref{fig3} is obtained as $ \omega_{L}=2.41499\times 10^{15} $ Hz.

Thirdly, using the second part of the red-detuned condition, i.e., $ \bar\Delta_{0}=\omega_d $ the optical mean-field or the intracavity photon number is determined by the relation $ n_{cav}=\bar a^{2}=\frac{\omega_{m}\Delta_0-\omega_{m}^{2}}{2(g_{0}^2+G_{0}^2)} $ where $ \Delta_0=\omega_{c}-\omega_{L}-N\frac{g_{a}^2}{2\Delta_a} $. For the black solid curve of Fig.~(\ref{fig3}) the intracavity photon number is obtained as $ n_{cav}\approx 2155 $. Now, based on the relation $ n_{cav}=\frac{E_{L}^2}{\kappa^{2}/4+\omega_{m}^{2}} $ the pump rate of the external driving laser is determined which for the black solid curve of Fig.~(\ref{fig3}) is obtained as $ E_L=1.899\times 10^8 $Hz which is consistent with the experimental data given in Refs.~\cite{RitterBECexp, BrennBECexp}. Similarly, for any other specified values of $ \mathcal{C}_0 $ and $ \mathcal{C}_1 $, one can calculate the frequency $\omega_{L}$ and the pump rate $E_L$ of the external driving laser which are necessary in an experimental setup to generate the theoretical results predicted in Figs.~(\ref{fig2}) and (\ref{fig3}). 

Here, we would like to compare the present parametrically driven hybrid optomechanical system with the parametrically driven bare one. For this purpose, let us compare the black solid curve of Fig.~(\ref{fig3}) with the red thick solid curve (representing the absence of BEC with $\mathcal{C}_0=0.04$ together with mechanical modulation) at zero temperature and at resonance frequency ($\omega\approx 0$). Based on Eq.~(\ref{sensitivity}) the sensitivity of the parametrically driven hybrid system (the black solid curve) is obtained as $\mathcal{S}|_{Hb}=5.82\times 10^{-20} \rm N/\sqrt{\rm Hz}$ while that of the parametrically driven bare system (red solid curve) is $\mathcal{S}|_{Br}=5.76\times 10^{-20} \rm N/\sqrt{\rm Hz}$. As is seen, both the (modulated) bare and the hybrid systems have nearly the same sensitivity. It should be noted that in order to have reliable results the SNR must be greater than a certain confident level, e.g., $r>3$ \cite{vitaliSNR}. For this purpose, either in the bare or in the hybrid system the input signal should be at least greater than $3\mathcal{S}$, i.e., $\tilde{F}>18\times 10^{-20} \rm N/\sqrt{\rm Hz}$ so that one can be assured that the signal has been detected in the system output.

Although, the sensitivity and the SNR of the present (modulated) hybrid system are nearly the same as those of (modulated) bare one, the (modulated) hybrid system has a much greater mechanical gain $(R_{m|Hb}\approx 118)$ in comparison to the bare one $(R_{m|Br}\approx 25)$ (about 5 times larger). Here, it should be noted that in the absence of modulations the signal is not amplified because $ R_{m|\rm {off-mods}} < 1 $. In order to see the advantage of an amplifier with a larger mechanical gain let us look at Eq.~(\ref{dPa}). As is seen, the role of the mechanical gain is that it amplifies both the input signal, i.e., $\tilde{F}(\omega)$ and the mechanical thermal noise, i.e., $\delta\hat X_b(\omega) $. It is because of the fact that the signal is entered into the system through the channel of the input thermal noise which is a normal phenomenon in most amplifiers. Since the mechanical gain makes the input signal as well as the input thermal noise be amplified simultaneously, the increase of the mechanical gain does not help the enhancement of the sensitivity and the SNR.

However, the important point is that if the signal is so weak that cannot be sensed by any instrument, one firstly needs a quantum instrument (a quantum detector/amplifier) that can sense the weak signal and also amplify it so much that it can be sensible for a classical electronic device that receives the output of the quantum amplifier and gives us a photocurrent which is equivalent to the output cavity spectrum. However, as has been already explained, the present amplifier (like others) amplifies both the signal and the thermal noise simultaneously. In this way, the SNR does not increase very much. Nevertheless, we will have an amplified signal (together with an amplified thermal noise) in the system output which is strong enough to be detected by an electronic device connected to the output of the amplifier.

Therefore, although the thermal noise in the system output has been simultaneously amplified but the important points are

i) The weak signal which was not previously detectable by the electronic device, has now been amplified so strongly that can be sensed by an electronic device which is connected to the output of our system.

ii) The output signal can be separated from the thermal noise (which is a white noise) by the well-known methods in electronics, especially if the signal frequency is known in advance. 
That is why the enhancement of the mechanical response is so important.

In short, our proposed hybrid system with both the atomic and mechanical modulations can act as a much better amplifier (because of its large mechanical gain) in comparison to the (modulated) bare optomechanical system which can amplify the input signal substantially while keeping the sensitivity nearly the same as that of the (modulated) bare one.\\

\section{summary and conclusion \label{summary}}

In this work, it has been proposed a scheme for an optomechanical force sensor composed of a hybrid optomechanical cavity containing an interacting cigar-shaped BEC where both the atomic collisions frequency of the BEC and the spring coefficient of the MO are coherently modulated. It has been shown that under these conditions the mechanical response of the system to the input signal is enhanced substantially which leads to the amplification of the weak input signal while the added noises of measurement can be maintained much below the SQL. In this way, such a hybrid system can operate as an ultra sensitive force sensor which can amplify the input signal without increasing the noise of measurement.

The advantage of the presented hybrid system in comparison to the bare optomechanical cavities is that the presence of the BEC together with atomic modulation improves the signal amplification substantially through the increase of the mechanical response of the system. Naturally, the price paid for this strong improvement of the signal amplification is an increase in the added noise of the measurement because the presence of the BEC, as an extra mode, induces an additional backaction noise. Nevertheless, the increment of the added noise due to the presence of the BEC is not so large to affect the measurement precision so that the sensitivity of the present (modulated) hybrid system  (as well as the SNR) is approximately of the same order of the (modulated) bare one ($\mathcal{S}|_{Hb,Br}\approx6\times 10^{-20} \rm N/\sqrt{\rm Hz}$). Nevertheless, the (modulated) hybrid system has a much greater mechanical gain in comparison to the (modulated) bare one.

Finally, it should be mentioned that the presented optomechanical force sensor has the optimum functionality near the on-resonance frequencies in the largely different cooperativities and red-detuned regimes where the impedance-matching condition is satisfied. Nevertheless, there exists the possibility of ultra precise measurement in the off-resonance region by controlling the BEC parameters and amplitudes of modulations which can enlarge the detection bandwidth.

\begin{acknowledgements} 
The authors would like to express their gratitude to the anonymous referee whose valuable comments and fruitful suggestions have improved the paper substantially. Furthermore, they would like to thank D.~Vitali, S.~H.~Tavassoli, M.~I.~Zibaii, S.~A.~Madani and B.~Levitan for the insightful discussions.
\end{acknowledgements}


%

\end{document}